\documentclass[epj]{svjour}
\usepackage{amsmath, graphicx, graphics}
\usepackage{dcolumn}
\usepackage{bm}

\begin{document}
\title{Electroweak phase transition in the economical 3-3-1 model}

\author{Vo Quoc  Phong\inst{1}\thanks{\emph{Present address:} vqphong@hcmus.edu.vn}, Hoang Ngoc Long\inst{2}\thanks{\emph{Present address:} hnlong@iop.vast.ac.vn}, Vo Thanh Van\inst{1}\thanks{\emph{Present address:} vtvan@hcmus.edu.vn},\and Le Hoang Minh\inst{1}\thanks{\emph{Present address:} lhminh167@gmail.com}}
  
\institute{Department of Theoretical Physics, Ho Chi Minh City University of Science, Vietnam \and Institute of Physics, Vietnamese  Academy of Science and Technology, 10 Dao Tan, Ba Dinh, Hanoi, Vietnam}

\date{Received: Received: date / Revised version: date}

\newcommand{\hs}{\hspace*{0.5cm}}
\newcommand{\vs}{\vspace*{0.5cm}}
\newcommand{\be}{\begin{equation}}
\newcommand{\ee}{\end{equation}}
\newcommand{\bea}{\begin{eqnarray}}
\newcommand{\eea}{\end{eqnarray}}
\newcommand{\ben}{\begin{enumerate}}
\newcommand{\een}{\end{enumerate}}
\newcommand{\bde}{\begin{widetext}}
\newcommand{\ede}{\end{widetext}}
\newcommand{\nn}{\nonumber}
\newcommand{\crn}{\nonumber \\}
\newcommand{\Tr}{\mathrm{Tr}}
\newcommand{\non}{\nonumber}
\newcommand{\noi}{\noindent}
\newcommand{\al}{\alpha}
\newcommand{\la}{\lambda}
\newcommand{\bet}{\beta}
\newcommand{\ga}{\gamma}
\newcommand{\va}{\varphi}
\newcommand{\om}{\omega}
\newcommand{\pa}{\partial}
\newcommand{\+}{\dagger}
\newcommand{\fr}{\frac}
\newcommand{\sq}{\sqrt}
\newcommand{\bc}{\begin{center}}
\newcommand{\ec}{\end{center}}
\newcommand{\Ga}{\Gamma}
\newcommand{\de}{\delta}
\newcommand{\De}{\Delta}
\newcommand{\ep}{\epsilon}
\newcommand{\varep}{\varepsilon}
\newcommand{\ka}{\kappa}
\newcommand{\La}{\Lambda}
\newcommand{\si}{\sigma}
\newcommand{\Si}{\Sigma}
\newcommand{\ta}{\tau}
\newcommand{\up}{\upsilon}
\newcommand{\Up}{\Upsilon}
\newcommand{\ze}{\zeta}
\newcommand{\ps}{\psi}
\newcommand{\Ps}{\Psi}
\newcommand{\ph}{\phi}
\newcommand{\vph}{\varphi}
\newcommand{\Ph}{\Phi}
\newcommand{\Om}{\Omega}

\abstract{We consider the EWPT in the economical 3-3-1 (E331) model. Our analysis shows that the EWPT in the model is a sequence of two first-order phase transitions, $SU(3) \rightarrow  SU(2)$ at the TeV scale and $SU(2) \rightarrow U(1)$ at the $100$ GeV scale. The EWPT $SU(3) \rightarrow SU(2)$ is triggered by the new bosons and the exotic quarks; its strength is about $1 - 13$ if the mass ranges of these new particles are $10^2 \,\mathrm{GeV} - 10^3 \,\mathrm{GeV}$. The EWPT $SU(2) \rightarrow U(1)$ is strengthened by only the new bosons; its strength is about $1 - 1.15$ if the mass parts of $H^0_1$, $H^\pm_2$ and $Y^\pm$ are in the ranges $10 \,\mathrm{GeV} - 10^2 \,\mathrm{GeV}$. The contributions of $H^0_1$ and $H^{\pm}_2$ to the strengths of both EWPTs may make them sufficiently strong to provide large deviations from thermal equilibrium and B violation necessary for baryogenesis.
\PACS{{11.15.Ex}{ Spontaneous breaking of gauge symmetries},
            {12.60.Fr}{ Extensions of electroweak Higgs sector} \and 
            {98.80.Cq}{ Particle-theory models (Early Universe)}}}
\authorrunning{Vo Quoc Phong, Hoang Ngoc Long, Vo Thanh Van}
\titlerunning{Electroweak phase transition in the economical 3-3-1 model}
\maketitle
\onecolumn
\section{INTRODUCTION}\label{secInt}

In the context of electroweak baryogenesis (EWBG), the EWPT plays an important role in explaining the Baryon Asymmetry of
  Universe (BAU) by electroweak physics. From the three Sakharov conditions, which are B violation, C and CP violations,
   and deviation from thermal equilibrium \cite{sakharov}, the EWPT should be a strongly first-order phase transition.
    That not only leads to thermal imbalance \cite{mkn}, but also makes a connection between B violation and CP
    violation via nonequilibrium physics \cite{ckn}.

The EWPT has been investigated in the Standard Model (SM) \cite{mkn,SME} as well as various extension models
\cite{BSM,SMS,dssm,munusm,majorana,thdm,ESMCO,lr,elptdm}. For the SM, although the EWPT strength is
 larger than unity at the electroweak scale, it is still too weak for the mass of the Higgs boson to be compatible with current
  experimental limits \cite{mkn,SME}; this suggests that EWBG requires new physics beyond the SM at the weak
   scale \cite{BSM}. Many extensions such as the Two-Higgs-Doublet model or Minimal Supersymmetric Standard Model
    have a more strongly first-order phase transition and the new sources of CP violation, which are necessary to account
     for the BAU; triggers for the first-order phase transition in these models are heavy bosons or dark matter
      candidates \cite{majorana,thdm,ESMCO,elptdm}.

Among the extensions beyond the SM, the models based on $\mathrm{SU}(3)_C\otimes \mathrm{SU}(3)_L \otimes \mathrm{U}(1)_X$
 gauge group (called 3-3-1 for short) \cite{ppf,flt} have some interesting features including the ability to explain the
 generation problem \cite{ppf,flt} and the electric charge quantization \cite{chargequan}. The structure of such a
  gauge group requires the 3-3-1 models to have at least two Higgs triplets. Thus the structure of symmetry breaking
  and the number of bosons are different from those in the SM.

In a previous work \cite{phonglongvan}, we have considered the EWPT in the reduced minimal 3-3-1 (RM331) model
 due to its simplicity, and found that our approach can be applied to the more complicated 3-3-1 models.
 In the present work, we follow the same approach for the economical 3-3-1 (E331) model \cite{ecn331}, whose lepton sector is more complicated than that of the RM331 model. The E331 model has the right-handed neutrino in the leptonic content, the bileptons (two singly charged gauge bosons $W^\pm$, $Y^\pm$, and a neutral  gauge bosons $X^0$), the heavy neutral
 boson $Z_2$, and the exotic quarks. The model has two Higgs triplets, and the physical scalar spectrum is composed of a singly
 charged scalar $H^\pm_2$ and a neutral scalars $H^0_1$ \cite{ecn331}. We will show in this paper that the new
  bosons and the exotic quarks can be triggers for the first-order phase transition in the model.

This paper is organized as follows. In Sec. \ref{sec2} we give a review of the E331 model on the Higgs, gauge boson, and lepton
 sectors. In Sec. \ref{sec3}, we find the effective potential in the model, which has a contribution from heavy bosons and exotic
 quarks as well as a contribution similar to that in the SM. In Sec. \ref{sec4}, we investigate the structure of the EWPT sequence
  in the E331 model, find the parameter ranges where the EWPTs are the strongly first-order to provide B violation necessary for
  baryogenesis, and show the constraints on the mass of the charged Higgs boson. Finally, we summarize and describe outlooks in Sec. \ref{sec5}.

\section{A REVIEW OF THE ECONOMICAL 331 MODEL}\label{sec2}

\subsection{Higgs potential}

In the E331 model, the 3-3-1 gauge group is spontaneously broken via two stages. In the first stage, the group
$\mathrm{SU}(3)_L \otimes \mathrm{U}(1)_X$
 breaks down to the $\mathrm{SU}(2)_L \otimes \mathrm{U}(1)_Y$ of the SM; and the second stage takes place as
  that we have known in the SM. This sequence of spontaneous symmetry breaking (SSB) is described by the Higgs potential \cite{ecn331}:
\bea  V(\chi,\phi)
 &=& \mu_1^2\chi^\dagger\chi
 + \mu_2^2 \phi^\dagger\phi
 + \lambda_1 (\chi^\dagger\chi)^2
 + \lambda_2 (\phi^\dagger\phi)^2+{}
 	\nonumber\\
 &{}&+ \lambda_3 (\chi^\dagger\chi)(\phi^\dagger\phi )
 + \lambda_4 (\chi^\dagger\phi)(\phi^\dagger\chi ),
\label{HiggsPotential}
\eea
in which $\chi$ and $\phi$ are the Higgs scalar triplets:
\be
\chi=
\begin{pmatrix}
\chi^{0}_1\\ \chi^-_2\\ \chi^{0}_3
\end{pmatrix}
\sim\left(1,3,-\frac{1}{3}\right)
, \hs
\phi=
\begin{pmatrix}
\phi^+_1\\ \phi^{0}_2\\ \phi^+_3
\end{pmatrix}
\sim\left(1,3,-\frac{1}{3}\right),
\label{higg}
\ee
whose VEVs are respectively given by:
\be
\langle\chi\rangle= \frac{1}{\sqrt{2}}
\begin{pmatrix}
u\\ 0\\ \omega
\end{pmatrix}, \qquad
\langle\phi\rangle= \frac{1}{\sqrt{2}}
\begin{pmatrix}
0\\ v\\ 0
\end{pmatrix},
\label{chankhong}
\ee
where the VEV $\om$ is responsible for the first stage, and the VEVs $u$ and $v$ are responsible for the second stage of
symmetry breaking. These VEVs satisfy the constraint \cite{ecn331}:
\be \label{constraint}
\omega \gg v\gg u.
\ee

The physical scalar spectrum of the model is composed of a charged scalar $H_2^{+}$, and two neutral scalars $H_1^0$ and
$H^0$. In this spectrum, $H^0$ is both the lightest neutral field and a $SU(2)_L$ component, hence it is identified as the
 SM Higgs boson. The Higgs content of the model can be summarized as follows:

\be
\chi = \left(\begin{array}{c} \fr{1}{\sqrt{2}}u+G_{X^0} \\ G_{Y^-}
\\ \fr{1}{\sqrt{2}}(\omega+H_1^0 + i G_{Z'})\end{array}\right)
, \hs
\phi = \left(\begin{array}{c}G_{W^+} \\
\fr{1}{\sqrt{2}} (v +H^0 +iG_Z) \\ H_2^+ \end{array}\right) \label{effHiggs},
\ee
where the Higgs masses are given by:
\bea
m^2_{H^0} & = &
\la_2 v^2+\la_1(u^2+\omega^2)-\sqrt{[ \la_2 v^2-\la_1(u^2+\omega^2)]^2+\la^2_3v^2(u^2+\omega^2)}\crn
&\approx& \frac{4\la_1\la_2-\la^2_3}{2\la_1}v^2,\label{mass-H0}\\
\hs
m^2_{H_1^0} &=& \la_2 v^2+\la_1(u^2+\omega^2)+\sqrt{[ \la_2 v^2-\la_1(u^2+\omega^2)]^2+\la^2_3v^2(u^2+\omega^2)} \crn
&\approx& 2\lambda_1\omega^2+\frac{\la^2_3}{2\la_1}v^2,\label{mass-H10}\\
m^2_{H_2^+}&=&\frac{\la_4}{2}(u^2+v^2+\omega^2).\label{mass-H2}
\eea

We note that in Ref. \cite{ecn331}, the mass formula of $H_1^0$ is approximate as $m^2_{H_1^0} \approx 2 \lambda_1\omega^2$.
 In the context of EWPT, however, we find that the better approximation should be that in Eq. \eqref{mass-H10}. Although the additional term $\frac{\la^2_3}{2\la_1}v^2$ is very small as compared to the first term and we may neglect it in some other considerations,
 it gives a very important contribution of $H_1^0$ to the EWPT $SU(2) \rightarrow U(1)$. 

\subsection{Gauge boson sector}\label{sectgauge}

The masses of the gauge bosons of this model come from the Lagrangian
\be \label{deriv}
\mathcal{L}^{GB}_{mass}=\left( \mathcal{D}_{\mu }\chi \right) ^{\dagger }\left( \mathcal{%
D}^{\mu }\chi \right) +\left( \mathcal{D}_{\mu }\phi \right)
^{\dagger }\left( \mathcal{D}^{\mu }\phi \right) ,
\ee where
\be \label{dhhp}
\mathcal{D}_\mu = \partial_\mu -ig T_iW_{i\mu} -ig_X T_9XB_{\mu},
\ee
with $T_9 =\frac{1}{\sqrt{6}} \textrm{diag}(1,1,1)$ so that $\textrm{Tr}(T_i T_j)=\delta_{ij}$.
The couplings of $SU(3)_L$ and $U(1)_X$ satisfy the relation:

\begin{equation}\label{coupling-SU3-U1}
t\equiv \fr{g_X}{g}= \fr{3\sq{2} s_W}{3-4 s_W^2},
\end{equation}
where $c_W=\cos\theta_W$, $s_W=\sin\theta_W$, $t_W=\tan \theta_W$, and $\theta_W$ is the Weinberg angle.

Eqs. \eqref{deriv} and \eqref{chankhong} lead to:
\bea \label{mw5}
W'^{\pm}_\mu&=&\frac{W_{1\mu} \mp iW_{2\mu}}{\sqrt{2}}
,\hs Y'^{\mp}_\mu=\frac{W_{6\mu} \mp iW_{7\mu}}{\sqrt{2}}
, \\ m_{W_5}^2&=&\fr{g^2}{4}(v^2+\om^2).
\eea

The combinations $W'$ and $Y'$ in \eqref{mw5} are mixed via a mass matrix:
\be \label{MassMatrix-charged}
\mathcal{L}^{CG}_{mass}=\frac{g^2}{4} (W'^-_\mu,Y'^-_\mu)
	\begin{pmatrix} u^2+v^2& u\om \\
		u\om& \om^2+v^2\end{pmatrix}
	\left(\begin{array}{c} 				
		W'^{+\mu}\\Y'^{-\mu}\end{array}\right).
\ee

Diagonalizing the mass matrix in Eq. \eqref{MassMatrix-charged}, we acquire the physical charged gauge bosons
\be W_\mu=\cos\theta\ W'_\mu-\sin\theta\ Y'_\mu
,\hs
Y_\mu=\sin\theta\ W'_\mu-\cos\theta\ Y'_\mu,
\ee
and their respective mass eigenvalues
\be \label{klwy}
m_W^2=\fr{g^2v^2}{4},\hs m_Y^2=\fr{g^2}{4}(u^2+v^2+\om^2),
\ee
where $\theta$ is the mixing angle which is defined by
\be \label{t_thetaDef}
t_\theta\equiv\tan\theta=\fr{u}{\om}.
\ee

The mass $m_W$ as in \eqref{klwy} suggests that the $W$ bosons of the model can be identified as those
 of the SM, and $v$ can be set as $v \simeq v_{weak}=246 \,\mathrm{GeV}$. From the constraints in
  \eqref{constraint}, $\theta$ should be very small, thus $W_\mu \simeq W'_\mu$ and $Y_\mu \simeq Y'_\mu$. Moreover,
the Michel parameter $\rho$ in the model connects $u$ with $v$ by the expression $\rho \approx 1
+\frac{3u^2}{v^2}$ \cite{ecn331}; and from the experimental data, $\rho=0.9987 \pm 0.0016$ \cite{pdg},
 that expression gives us $\frac{u}{v}\leq 0.01$, which leads to $u<2.46 \,\mathrm{GeV}$.  With $\om$ in the
  range $1 \,\mathrm{TeV} - 5 \,\mathrm{TeV}$, we have
\be \label{t_theta}
t_\theta=\frac{u}{\om} \approx 0.001.
\ee

For the neutral gauge bosons, the mass matrix in the basis ($W_{3\mu}, W_{8\mu}, B_\mu, W_{4\mu}$) is given by
\begin{equation}\label{MassMatrix-neutral}
M^2 = \fr{g^2}{4}\left(%
\begin{array}{cccc}
  u^2+v^2 & \fr{u^2-v^2}{\sq{3}}
  &-\fr{2t}{3\sq{6}}(u^2+2v^2) & 2u\om
\\\fr{u^2-v^2}{\sq{3}} & \fr 1 3 (4\om^2+u^2 +v^2) &
  \fr{\sq{2}t}{9}(2\om^2 -u^2 +2v^2) & -\fr{2}{\sq{3}} u\om
\\-\fr{2t}{3\sq{6}}(u^2+2v^2) & \fr{2t}{9}(2\om^2 -u^2 +2v^2) & -\fr{2t^2}{27}(\om^2+u^2+4v^2) & -\fr{8t}{3\sq{6}}u\om
\\2u\om & -\fr{2}{\sq{3}} u\om & -\fr{8t}{3\sq{6}}u\om & u^2+\om^2
\end{array}%
\right).
\end{equation}
The diagonalization of the mass matrix in Eq. \eqref{MassMatrix-neutral} leads to the mass eigenstates of four following neutral gauge bosons:
\begin{eqnarray}
m^2_\ga\ &=& 0,\hs
m^2_{W'_{4}}=\fr{g^2}{4}(u^2+\om^2) \label{mw4}  ,\\
 m^2_{Z_1}&=& [2g^{-2} \sq{3-4s_W^2}]^{-1}
 	\left\lbrace
 		[ c_W^2(u^2+\om^2)+v^2
 	\right.\crn
 	&&\left.-\sq{
 		 	[c_W^2(u^2+\om^2)+v^2]^2
 			+(3-4s_W^2)(3u^2\om^2-u^2v^2-v^2\om^2)
 		}
 	\right\rbrace
 ,\\
 m^2_{Z_2}&=&	[2g^{-2} \sq{3-4s_W^2}]^{-1}
 	\left\lbrace
 		[ c_W^2(u^2+\om^2)+v^2
 	\right.\crn
 	&&\left.+\sq{
 		 	[c_W^2(u^2+\om^2)+v^2]^2
 			+(3-4s_W^2)(3u^2\om^2-u^2v^2-v^2\om^2)
 		}
 	\right\rbrace.
\end{eqnarray}

Due to the constraints \eqref{constraint}, the physical states $Z_1$ and $Z_2$ get masses
\be \label{klz1z2}
m^2_{Z_1}=\fr{g^{2}}{4c_W^2} (v^2-3u^2),\hs
m^2_{Z_2}=\fr{g^2c_W^2\om^2}{3-4s_W^2}.
\ee

Since the components $W'_4$ and $W_5$ have the same mass, we can identify their  combination,
\be \label{X}
X^0_\mu=\frac{1}{\sq{2}}
	(W'_{4\mu} - iW_{5\mu}),
\ee
as a physical neutral non-Hermitian gauge boson, which carries the lepton number with two units. The subscript
$0$ of $X_\mu$ in Eq. \eqref{X} denotes neutrality of the gauge boson $X$ but sometimes this subscript may be dropped.

\subsection{Fermion sector}

The fermion content in this model, which is anomaly free, is given by
\bea
\psi_{iL}&=&\begin{pmatrix}
\nu_i\\ e_i\\ \chi^{0}_i
\end{pmatrix}_L \sim \left(1,3,-\fr{1}{3}\right)
,\hs e_{iR}\sim(1,1,-1),\hs i=1,2,3, \crn
Q_{1L}&=&\begin{pmatrix}
u_1\\ d_1\\ U
\end{pmatrix}_L \sim \left(3,3,\fr{1}{3}\right)
,\hs
Q_{\alpha L}=\begin{pmatrix}
d_\alpha\\ u_\alpha\\ D_\alpha
\end{pmatrix}_L \sim \left(3,3^*,0\right)
,\hs \alpha=2,3, \crn
u_{iR}&\sim&\left(3,1,\fr{2}{3}\right),\hs
d_{iR}\sim\left(3,1,-\fr{1}{3}\right),\hs
u_{R}\sim\left(3,1,\fr{2}{3}\right),\hs
D_{\alpha R}\sim\left(3,1,-\fr{1}{3}\right).
\eea

The Yukawa interactions which induce masses for the fermions can be written as
\be
\mathcal{L}_{Yuk} = \mathcal{L}_{LNC} + \mathcal{L}_{LNV} \label{exoticquarks}
\ee
in which $\mathcal{L}_{LNC}$ is the Lagrangian part for lepton number conservation and $\mathcal{L}_{LNV}$ is
 that for lepton number violation. These Lagrangian parts are given by:
\bea
\mathcal{L}_{LNC}&=&
h^U \bar{Q}_{1L} \chi U_{R}
+h^D_{\alpha\beta}\bar{Q}_{\alpha L}
	\chi^*
	D_{\beta R} \crn
&&+h^e_{ij}\bar{\psi}_{iL}\phi e_{jR}
+h^\epsilon_{ij}	
	\epsilon_{abc} (\bar{\psi}^c_{iL})_a
	(\psi^c_{jL})_b
	(\phi)_c \crn
&&+h^d_{i} \bar Q_{1L} \phi d_{iR}
+h^u_{\alpha i} \bar Q_{\alpha L}\phi^* u_{iR}
+ H.c.
\crn
\mathcal{L}_{LNV}&=&
s^u_{i}\bar Q_{1L}\chi u_{iR}
+s^d_{\alpha i}\bar{Q}_{\alpha L}
	\chi^*
	d_{iR} \crn
&&+s^D_{\alpha} \bar Q_{1L} \phi D_{\alpha R}
+s^U_{\alpha} \bar{Q}_{\alpha L}
	\phi^*
	U_{R}
+ H.c.
\eea
where a, b and c stand for the $SU(3)_L$ indices.

During the SSB sequence of this model, the VEV $\om$ gives the masses for the exotic quarks $U$ and $D_\alpha$,
the VEV $u$ which is the source of lepton-number violations gives the masses for the quarks $ u_1$ and $d_\alpha$,
 the VEV $v$ gives the masses for the quarks $u_\alpha$ and $d_1$ as well as all ordinary leptons.

\section{EFFECTIVE POTENTIAL IN THE ECONOMICAL 331 MODEL} \label{sec3}

From the Higgs potential \eqref{HiggsPotential}, we obtain $V_{0}$  in a form which is dependent on the VEVs as follows:
\begin{equation}\label{V0}
V_{0}(u,\om,v)= \frac{\mu^2_1}{2} (u^2+\omega^2)
+\frac{\mu^2_2}{2} v^2
+\frac{\lambda_1}{4}(u^4+\omega^4+2u^2\omega^2)
+\frac{\lambda_2}{4}v^4+\frac{\lambda_3}{4}(u^2 v^2+v^2 \omega^2).
\end{equation}

We see that $V_{0}(u,\om,v)$ has a quartic form like in the SM, but it depends on three variables, $u$, $\om$ and $v$; it also has the mixings between these variables. However, we can transform $u$ into $\om$ by $t_{\theta}$ as defined in Eq. \eqref{t_thetaDef}. We note that, if the Universe' energies allow of the existence of the gauge symmetry $ \mathrm{SU}(3)_L \otimes \mathrm{U}(1)_X$ and the SSB sequence in the E331 model, the VEVs $u$, $\om$ and $v$ must satisfy the constraint \eqref{constraint}. This leads to $t_{\theta} \ll 1$, and we can neglect the contribution of $u$. On the other hand, by developing the Higgs potential (\ref{HiggsPotential}), we obtain two minimum equations which permit us to transform the mixing between $\om$ and $v$,

\bea
\mu^2_1+\lambda_1(u^2+\omega^2)+\lambda_3\frac{v^2}{2}&=&0,\crn
\mu^2_2+\lambda_2v^2+\lambda_3\frac{(u^2+\omega^2)}{2}&=&0.
\label{eqn17Feb}
\eea
From Eq.(\ref{eqn17Feb})  we obtain
\begin{equation}
\lambda_3(u^2+\omega^2)v^2=-2(\mu^2_2v^2+\lambda_2v^4),\label{c}
\end{equation}
and
\begin{equation}
u^2\omega^2=\frac{\mu^2_2-\frac{2\lambda_2}{\lambda_3}\mu^2_1}{\frac{2\lambda_1\lambda_2}{\lambda_3}-\frac{\lambda_3}{2}}u^2-u^4.\label{d}
\end{equation}
Substituting    Eqs.(\ref{c}) and (\ref{d}) into  Eq.(\ref{V0})  yields
\begin{equation}\label{V01}
V_{0}(u,\om,v)
=\frac{\mu^2_1}{2}\omega^2+\frac{\lambda_1}{4}\omega^4+\left[\frac{\mu^2_1}{2}+2\frac{\mu^2_2-\frac{2\lambda_2}{\lambda_3}\mu^2_1}{\frac{2\lambda_1\lambda_2}{\lambda_3}-\frac{\lambda_3}{2}}\right]u^2-\frac{\lambda_1}{4}u^4-\frac{\lambda_2v^4}{4}.
\end{equation}

Neglecting  $u$ and from those relations, we can write $V_0$ in Eq. \eqref{V01} as a sum of two parts corresponding to two stages of SSB:
\begin{equation}\label{V0Split}
V_0(\om,v)=V_0(\om)+V_0(v),
\end{equation}
where  $V_0(\om)=\frac{\mu^2_1}{2}\omega^2+\frac{\lambda_1}{4}\omega^4$ and $V_0(v)=-\frac{\lambda_2v^4}{4}$ are in the quartic form. In addition, we have alternative  ways to arrive Eq. \eqref{V0Split} which has other forms but $V_0(\om)$ and $V_0(v)$ are still in the quartic form.

In order to derive the effective potential, we start from the full Higgs Lagrangian:
\begin{equation} \label{HiggsLagrangian}
\mathcal{L}=\mathcal{L}^{GB}_{mass}+V(\chi,\phi),
\end{equation}
where $\mathcal{L}^{GB}_{mass}$ and $V(\chi,\phi)$ are respectively given by Eq. \eqref{deriv} and Eq. \eqref{HiggsPotential}.

In order to see the effective potential of this model can be split into two separated parts, we analyze the processes which generate the masses for all particles. The first, we want to mention the masses of the gauge fields which come from Eq. (\ref{deriv}), we can rewrite this equation as form:
\be \label{deriv2}
\mathcal{L}^{GB}_{mass}=\left( \mathcal{D}_{\mu }\langle \chi\rangle \right) ^{\dagger }\left( \mathcal{%
D}^{\mu }\langle \chi\rangle \right) +\left( \mathcal{D}_{\mu }\langle \phi\rangle \right)
^{\dagger }\left( \mathcal{D}^{\mu }\langle \phi\rangle \right) =  A+B,
\ee 
where
\bea
A &\equiv & \left( \mathcal{D}_{\mu }\langle \chi\rangle \right) ^{\dagger }\left( \mathcal{%
D}^{\mu }\langle \chi\rangle \right)
,\crn
B&\equiv&\left( \mathcal{D}_{\mu }\langle \phi\rangle \right)
^{\dagger }\left( \mathcal{D}_{\mu }\langle \phi\rangle \right).
\nn
\eea
Note that the gauge fields ($W^\mu_i, B^\mu$) inside the covariant derivatives of the $A$ and $B$  are the same. So after diagonalizing, the gauge fields in the $A$ and $B$  are the same, and we obtain gauge bosons: $\ga $, $Z_1$, $Z_2$, $X_0$, $W^{\pm}$, $Y^{\pm}$.

Neglecting $u$ and from the term $A$, one obtains the mass components of the physical gauge bosons only depend on $\omega$ and $u$ (for details see, the second paper in \cite{ecn331}):
\begin{equation}\label{ma}
\begin{split}
M^A_{bosons}=&m^2_{W^{\pm}}(\omega, u)W^{+}_{\mu}W^{-\mu}+m^2_{Y^{\pm}}(\omega,u)Y^{+}_{\mu}Y^{-\mu}\\
&+m^2_{X^0}(\omega,u)X^0_{\mu}Z^{0\mu}+m^2_{Z^1}(\omega,u)Z^1_{\mu}Z^{1 \mu}+m^2_{Z^2}(\omega,u)Z^2_{\mu}Z^{2 \mu}.
\end{split}
\end{equation}

From the term $B$, one obtains the mass components of the physical gauge bosons only depend on $v$:
\begin{equation}\label{mb}
\begin{split}
M^B_{bosons}=&m^2_{W^{\pm}}(v)W^{+}_{\mu}W^{-\mu}+m^2_{Y^{\pm}}(v)Y^{+}_{\mu}Y^{-\mu}\\
&+m^2_{X^0}(v)X^0_{\mu}Z^{0\mu}+m^2_{Z^1}(v)Z^1_{\mu}Z^{1 \mu}+m^2_{Z^2}(v)Z^2_{\mu}Z^{2 \mu}.
\end{split}
\end{equation}

Looking carefully at Eq. (\ref{MassMatrix-charged}) and (\ref{MassMatrix-neutral}), this is   easily checked. 
So we obtain $m^2_{W^{\pm}}(\omega, u)=m^2_{X^0}(v)=0$, $m^2_{Z_1}(\omega,u)=m^2_{Z_2}(v)\approx 0$. 

Through this analysis we see that when we combine A with B to make the diagonalization of the matrix in the above section, this leads to a mixing between the gauge bosons in the $A$ and $B$. However we note that although the fields of the $A$ and $B$ are the same, the fields in the $A$ or Eq. (\ref{ma}) only go with $\om$ and $u$, the fields in the $B$ or Eq. (\ref{mb}) only go with $v$. Because the $A$ and $B$ are not mix together. Therefore we found that the physical gauge bosons are like "break", this problem is due to the initial assumptions of the model about the covariant derivative.

The second, we want to mention the masses of the Higgs fields. The Higgs potential, Eq. (\ref{HiggsPotential}), has two last components, $\lambda_3 (\chi^\dagger\chi)(\phi^\dagger\phi )$ and $\lambda_4 (\chi^\dagger\phi)(\phi^\dagger\chi )$, so we have a mixing among VEVs. In the calculation in the above section, we perform an approximation in Eqs. (\ref{mass-H0}), (\ref{mass-H10}) and (\ref{mass-H2}). These approximations  did lose the mixing between VEVs, so they did lose $\lambda_3 (\chi^\dagger\chi)(\phi^\dagger\phi )$ and $\lambda_4 (\chi^\dagger\phi)(\phi^\dagger\chi )$ or these two last components are absorbed into the other components of the Higgs potential.

Therefore the masses of the gauge bosons and the Higgses presented in Table \ref{tab1}, from which we can split the boson masses into two parts for two SSB stages:
\begin{equation}\label{BosonMassSplit}
m^2_{boson}(\om, v)= m^2_{boson}(\om)+m^2_{boson}(v).
\end{equation}

However,  note that the Higgs fields are like the gauge bosons. A field function multiplies  by the square of the mass that contains a VEV, so that the field only effects on the VeV, but it does not effect on all of VEVs.

\begin{table}
\caption{Mass formulations of bosons in the E331 model}
\bc
\begin{tabular}{|l|l|c|c|c|}
\hline
\hline
 Bosons& $m^2(\om,v)$& $ m^2(\om)$ &$m^2(v)$& $m^2(v_0=246 GeV)$ \\
\hline
$m_{W^{\pm }}^{2}$& $\frac{g^{2}}{4}v^{2}$&
0& $\frac{g^{2}}{4}v^{2}$ &$80.39^2$ $(\mathrm{GeV})^2$ \\
\hline
$m_{Y^{\pm }}^{2}$& $\frac{g^{2}}{4}(\om^2+v^{2})$&
$\frac{g^{2}}{4}\om^2$ & $\frac{g^{2}}{4}v^{2}$& $80.39^2$ $(\mathrm{GeV})^2$\\
\hline
$m_{X^0}^{2}$& $\frac{g^{2}}{4}\om^2$&
$\frac{g^{2}}{4}\om^2$& 0 &0 \\
\hline
$m^2_{Z_1}\sim m^2_{Z}$& $\frac{g^2}{4c_W^2} v^2$&
0&$\frac{g^2}{4c_W^2} v^2$& $91.68^2$ $(\mathrm{GeV})^2$ \\
\hline
$m^2_{Z_2}\sim m^2_{Z'}$& $\frac{g^2c_W^2}{3-4s_W^2}\om^2$
&$\frac{g^2c_W^2}{3-4s_W^2}\om^2$& 0&0 \\
\hline
$m^2_{H^0}$& $\left(2\la_2 -\frac{\la^2_3}{2\la_1}\right)v^2$&
0& $\left(2\la_2 -\frac{\la^2_3}{2\la_1}\right)v^2$& $125^2$ $(\mathrm{GeV})^2$\\
\hline
$m^2_{H^0_1}$& $2\lambda_1\om^2+\frac{\la^2_3}{2\la_1}v^2$ &
$2\lambda_1\om^2$& $\frac{\la^2_3}{2\la_1}v^2$ &$\frac{\la^2_3}{2\la_1}v_0^2$\\
\hline
$m^2_{H^\pm_2}$& $\fr{\la_4}{2}(\om^2+v^2)$&
$\fr{\la_4}{2}\om^2$ & $\fr{\la_4}{2}v^2$&$\fr{\la_4}{2}v_0^2$ \\
\hline
\hline
\end{tabular}
\ec
\label{tab1}
\end{table}

The last, expanding the Higgs fields $\chi$ and $\phi$ around their VEVs which are $u$, $\om$ and $v$, we obtain
\bea \label{HiggsLagrangian-1}
\begin{split}
\mathcal{L}
=&\frac{1}{2}\partial^{\mu}\om\partial_{\mu}\om+
\frac{1}{2}\partial^{\mu}v\partial_{\mu}v+V_0(\om,v)
+M^A_{bosons}+M^B_{bosons}\\
&+\sum m_{exotic-quarks}(\om)Q\bar{Q}+m_{top-quark}(v)t\bar{t},
\end{split}
\eea
In the E331 model, we have two massive bosons like the SM bosons $Z_1$ and  $W^{\pm}$, two new heavy neutral bosons $X_0$ and $Z_2$, the singly charged gauge bosons $Y^{\pm}$, one singly charged Higgs $H^\pm_2$, one heavy neutral Higgs $H^0_1$ and one SM-like Higgs $H^0$.  We must consider contributions from all fermions and bosons. But for fermions, we retain only the top and exotic quarks because their contributions dominate over those from the other fermions \cite{mkn}. Therefore, from the Lagrangain \eqref{HiggsLagrangian-1} we acquire two motion equations according to $\om$ and $v$,
\bea
 \partial^{\mu}\om\partial_{\mu}\om +\frac{\partial V_0(\om)}{\partial
\om}+\sum \frac{\partial m^2_{bosons}(\om)}{\partial \om}W^{\mu}
W_{\mu}+\sum\frac{\partial m_{exotic-quarks}(\om)}{\partial \om}Q\bar{Q} &= &0,\label{mchi1}\\
\partial^{\mu}v\partial_{\mu}v+\frac{\partial V_0(v)}{\partial v}
+\sum \frac{\partial m^2_{bosons}(v)}{\partial v}W^{\mu}W_{\mu}+
\frac{\partial m_{top-quark}(v)}{\partial v}t\bar{t}&=& 0, \label{mrho1}
\eea
\textbf{where $W$ runs over all gauge fields and Higgs bosons. From Eq. \eqref{mchi}, averaging over space we obtain}
\bea
 \partial^{\mu}\om\partial_{\mu}\om +\frac{\partial V_0(\om)}{\partial
\om}+\sum \frac{\partial m^2_{bosons}(\om)}{\partial \om}\left\langle W^{\mu}
W_{\mu}\right\rangle+\sum\frac{\partial m_{exotic-quarks}(\om)}{\partial \om}\left\langle Q\bar{Q}\right\rangle &= &0,\label{mchi}\\
\partial^{\mu}v\partial_{\mu}v+\frac{\partial V_0(v)}{\partial v}
+\sum \frac{\partial m^2_{bosons}(v)}{\partial v}\left\langle W^{\mu}W_{\mu}\right\rangle+
\frac{\partial m_{top-quark}(v)}{\partial v}\left\langle t\bar{t}\right\rangle&=& 0. \label{mrho}
\eea

Note that $\left\langle W^{\mu}W_{\mu}\right\rangle$ in Eq. (\ref{mchi}) only effect on $\om$, so it only depends on $m_{bosons}(\om)$. Similarly, $\left\langle W^{\mu}W_{\mu}\right\rangle$ in Eq. (\ref{mrho}) only effect on $v$, so it only depends on $m_{bosons}(v)$. 

Using Bose-Einstein and Fermi-Dirac distributions respectively for bosons and fermions to average over space, we obtain the one-loop effective potential $V_{eff}(\om)$ for the electroweak phase transition $SU(3)-SU(2)$ at high temperatures:

\bea \label{EP-o}
V_{eff}(\om)&=& V_{0}(\om)
+\frac{1}{64\pi^2}\left[
	6m^4_Y(\om)\ln\frac{m^2_Y(\om)}{Q'^2}
	+6m^4_X(\om)\ln\frac{m^2_X(\om)}{Q'^2} \right.\crn
&&	+3m^4_{Z_2}(\om)\ln\frac{m^2_{Z_2}(\om)}{Q'^2}	
	+m^4_{H^0_1}(\om)\ln\frac{m^2_{H^0_1}(\om)}{Q'^2} \crn
&&\left.			
	+2m^4_{H^+_2}(\om)\ln\frac{m^2_{H^+_2}(\om)}{Q'^2}
	-36m^4_Q(\om)\ln\frac{m^2_Q(\om)}{Q'^2}\right]\crn
&+&\frac{T^4}{4\pi^2}\left[
	6F_{-}\left(\frac{m_Y(\om)}{T}\right) +6F_{-}\left(\frac{m_X(\om)}{T}\right)
	+3F_{-}\left(\frac{m_{Z_2}(\om)}{T}\right) \right. \crn
&&\left.	
	+F_{-}\left(\frac{m_{H^0_1}(\om)}{T}\right)
	+2F_{-}\left(\frac{m_{H^+_2}(\om)}{T}\right)
	+36F_{+}\left(\frac{m_Q(\om)}{T}\right)
	\right],
\eea
in which $m_Q$ indicates the masses of three exotic quarks. Similarly, from Eq. \eqref{mrho}, we obtain the high-temperature effective potential $V_{eff}(v)$ for the electroweak phase transition $SU(2)-U(1)$:
\bea \label{EP-v}
V_{eff}(v)&=& V_{0}(v)
+\frac{1}{64\pi^2}\left[
	6m^4_W(v)\ln\frac{m^2_W(v)}{Q^2}
	+6m^4_Y(v)\ln\frac{m^2_Y(v)}{Q^2}
	\right.\crn
&&	+3m^4_{Z_1}(v)\ln\frac{m^2_{Z_1}(v)}{Q^2}
	+m^4_{H^0}(v)\ln\frac{m^2_{H^0}(v)}{Q^2}+m^4_{H^0_1}(v)\ln\frac{m^2_{H^0_1}(v)}{Q^2} \crn
&&\left.
	+2m^4_{H_2^+}(v)\ln\frac{m^2_{H_2^+}(v)}{Q^2}
	-12m^4_t(v)\ln\frac{m^2_t(v)}{Q^2}\right]\crn
&+&\frac{T^4}{4\pi^2}\left[	
	6F_{-}\left(\frac{m_W(v)}{T}\right)
	+6F_{-}\left(\frac{m_Y(v)}{T}\right)
	+3F_{-}\left(\frac{m_{Z_1}(v)}{T}\right)
	\right.\crn
&&\left.	
	+F_{-}\left(\frac{m_{H^0}(v)}{T}\right)+F_{-}\left(\frac{m_{H^0_1}(v)}{T}\right)
	+2F_{-}\left(\frac{m_{H^+_2}(v)}{T}\right)
	+12F_{+}\left(\frac{m_t(v)}{T}\right)
	\right],
\eea
in which $m_t$ indicates the mass of the top quark. $F_{\mp}\left(\frac{m}{T}\right)$ come from $\left\langle W^{\mu}W_{\mu}\right\rangle$ and describe the thermal contributions of particles with masses $m$. These terms are given by
\bea
F_{\mp}\left(\frac{m}{T}\right)&=&\int^{\frac{m}{T}}_0\alpha
J^{(1)}_{\mp}(\alpha,0)d\alpha, \label{F}
\eea
where 
\bea
J^{(1)}_{\mp}(\alpha,0)&=&2\int^{\infty}_{\alpha}\frac{(x^2-\alpha^2)^{1/2}}{e^{x}\mp
1}dx. \label{J}
\eea

As the above analyze, $F_{\mp}\left(\frac{m}{T}\right)$ in Eq. (\ref{EP-o}) only depend on $\om$ and $F_{\mp}\left(\frac{m}{T}\right)$ in Eq. (\ref{EP-v}) only depend on $v$. 
 
Eqs. \eqref{HiggsLagrangian-1}-\eqref{EP-o} and \eqref{EP-v} do not consist of any mixing between $\om$ and $v$. Therefore, we can write the total effective potential in the E331 model as
\begin{equation}\label{EP-E331}
V^{E331}_{eff}=V_{eff}(\om)+V_{eff}(v).
\end{equation}

The effective potentials $V_{eff}(\om)$ and $V_{eff}(v)$ seem to depend on the arbitrary scales $Q'$ and $Q$ respectively. However, by the same reasoning as in \cite{mkn}, we can show that the structure of these potentials remain unchanged for the changes in scales. At zero temperarure, all thermal contributions vanish, and due to the quartic form of $V_0(\om)$ and $V_0(v)$, we can rewrite Eqs. \eqref{EP-o} and \eqref{EP-v} as
\bea \label{EP-o-0K-1}
V^{0^o K}_{eff}(\om)&=& \lambda'_R \om^4+M'^2_R\om^2+\Lambda'_R
+\frac{1}{64\pi^2}\left[
	6m^4_Y(\om)\ln\frac{m^2_Y(\om)}{Q'^2}
	+6m^4_X(\om)\ln\frac{m^2_X(\om)}{Q'^2} \right.\crn
&&	+3m^4_{Z_2}(\om)\ln\frac{m^2_{Z_2}(\om)}{Q'^2}	
	+m^4_{H^0_1}(\om)\ln\frac{m^2_{H^0_1}(\om)}{Q'^2} \crn
&&\left.			
	+2m^4_{H^+_2}(\om)\ln\frac{m^2_{H^+_2}(\om)}{Q'^2}
	-36m^4_Q(\om)\ln\frac{m^2_Q(\om)}{Q'^2}\right],
	\eea
and
\bea \label{EP-v-0K-1}
V^{0^o K}_{eff}(v)&=& \lambda_R v^4+M^2_R v^2+\Lambda_R
+\frac{1}{64\pi^2}\left[
	6m^4_W(v)\ln\frac{m^2_W(v)}{Q^2}
	+6m^4_Y(v)\ln\frac{m^2_Y(v)}{Q^2}
	\right.\crn
&&	+3m^4_{Z_1}(v)\ln\frac{m^2_{Z_1}(v)}{Q^2}
	+m^4_{H^0}(v)\ln\frac{m^2_{H^0}(v)}{Q^2}+m^4_{H^0_1}(v)\ln\frac{m^2_{H^0_1}(v)}{Q^2} \crn
&&\left.
	+2m^4_{H_2^+}(v)\ln\frac{m^2_{H_2^+}(v)}{Q^2}
	-12m^4_t(v)\ln\frac{m^2_t(v)}{Q^2}\right],
\eea
where $\lambda'_R$, $M'_R$, $\Lambda'_R$, $\lambda_R$, $M_R$, and $\Lambda_R$ are the renormalized constants. The changes such as $Q' \to \kappa' Q'$ (or $Q \to \kappa Q$) induce the terms which contain $\kappa'$ (or $\kappa$) and are proportional to $m^4_{boson}(\om) \sim \om^4$ (or $m^4_{boson}(v) \sim v^4$). Those terms can be absorbed by $\lambda'_R$ (or $\lambda_R$). This makes the physics remain the same.

By this reason, we can put $Q'=\epsilon' \om_0$ and $Q=\epsilon v_0$ into Eqs. \eqref{EP-o-0K-1} and \eqref{EP-v-0K-1}, respectively. Combining the terms which contain $\epsilon'$ and $\epsilon$ with the renormalized constants, we have: 
\bea \label{EP-o-0K-2}
V^{0^o K}_{eff}(\om)&=& \frac{\lambda'_0}{4} \om^4+M'^2_0\om^2+\Lambda'_0
+\frac{1}{64\pi^2}\left[
	6m^4_Y(\om)\ln\frac{\om^2}{\om_0^2}
	+6m^4_X(\om)\ln\frac{\om^2}{\om_0^2} \right.\crn
&&	+3m^4_{Z_2}(\om)\ln\frac{\om^2}{\om_0^2}	
	+m^4_{H^0_1}(\om)\ln\frac{\om^2}{\om_0^2} \crn
&&\left.			
	+2m^4_{H^+_2}(\om)\ln\frac{\om^2}{\om_0^2}
	-36m^4_Q(\om)\ln\frac{\om^2}{\om_0^2}\right],
	\eea
and
\bea \label{EP-v-0K-2}
V^{0^o K}_{eff}(v)&=& \frac{\lambda_0}{4} v^4+M^2_0 v^2+\Lambda_0
+\frac{1}{64\pi^2}\left[
	6m^4_W(v)\ln\frac{v^2}{v_0^2}
	+6m^4_Y(v)\ln\frac{v^2}{v_0^2}
	\right.\crn
&&	+3m^4_{Z_1}(v)\ln\frac{v^2}{v_0^2}
	+m^4_{H^0}(v)\ln\frac{v^2}{(v_0)^2}+m^4_{H^0_1}(v)\ln\frac{v^2}{v_0^2} \crn
&&\left.
	+2m^4_{H_2^+}(v)\ln\frac{v^2}{v_0^2}
	-12m^4_t(v)\ln\frac{v^2}{v_0^2}\right],
\eea
where $\lambda'_0$, $M'^2_0 $, $\Lambda'_0$, $\lambda_0$, $M^2_0$, $\Lambda_0$ are the parameters those can be specified from the conditions \eqref{MinCondition-o} and \eqref{MinCondition-v}. And we acquire:
\bea
\lambda'_0 &=&\left\{
 \frac{m_{H^0_1}^2(\om_0)}{2 \om_0^2}
 	- \frac{3}{32\pi^2} \left(
 		6 m_Y^4(\om_0) +6 m_X^4(\om_0) +3 m_{Z_2}^4(\om_0)
     \right.\right.	\crn
	&&\qquad \left.\left. 		
		+m_{H^0_1}^4(\om_0) +2m_{H^+_2}^4(\om_0)-36 m_Q^4(\om_0)
		\right)\right\},\crn 
\\
{M'}_0^2 &=& \left\{
	-\frac{1}{4} m_{H^0_1}^2(\om_0) +
 	\frac{1}{32\pi^2\om_0^2} \left(
 		6 m_Y^4(\om_0) +6 m_X^4(\om_0) +3 m_{Z_2}^4(\om_0)
 		\right.\right.	\crn
	&&\qquad \left.\left.
		+m_{H^0_1}^4(\om_0) +2m_{H^+_2}^4(\om_0)-36 m_Q^4(\om_0)
		\right)\right\}	,\crn
\\
{\Lambda'_0}&=&\frac{\om_0^2}{4}\left\{\frac{m_{H^0_1}^2(\om_0)}{2}-\frac{1}{32\pi^2\om_0^2} \left(
 		6 m_Y^4(\om_0) +6 m_X^4(\om_0) +3 m_{Z_2}^4(\om_0)
 		\right.\right.	\crn
	&&\qquad \left.\left.
		+m_{H^0_1}^4(\om_0) +2m_{H^+_2}^4(\om_0)-36 m_Q^4(\om_0)
		\right)\right\};
\eea
\bea
\lambda_0 &=&\left\{
 \frac{m_{H^0}^2(v_0)+m_{H^0_1}^2(v_0)}{2 v_0^2}
 	- \frac{3}{32\pi^2} \left(
		6 m_W^4(v_0) +6 m_Y^4(v_0) +3m_{Z_1}^4(v_0)
\right.\right.	\crn
	&&\qquad \left.\left. 
	+m_{H^0}^4(v_0)+m_{H^0_1}^4(v_0) +2m_{H^+_2}^4(v_0) -12 m_t^4 (v_0)
	\right)\right\},\crn 
\\
M_0^2 &=& \left\{-\frac{m_{H}^2(v_0)+m_{H^0_1}(v_0)}{4}
	+\frac{1}{32\pi^2v_0^2} \left(
		6 m_W^4(v_0) +6 m_Y^4(v_0) +3m_{Z_1}^4(v_0)
	\right.\right.\crn
	&&\left.\left.\qquad
		+m_{H^0}^4(v_0)+m_{H^0_1}^4(v_0) +2m_{H^+_2}^4(v_0) -12 m_t^4 (v_0)
	\right)\right\},\crn
\\
\Lambda_0&=&\frac{v^2_0}{4}\left\{\frac{m_{H^0}^2(v_0)+m_{H^0_1}^2(v_0)}{2}-\frac{1}{32\pi^2 v_0^2} \left(
		6 m_W^4(v_0) +6 m_Y^4(v_0) +3m_{Z_1}^4(v_0)
\right.\right.	\crn
	&&\qquad \left.\left. 
	+m_{H^0}^4(v_0)+m_{H^0_1}^4(v_0) +2m_{H^+_2}^4(v_0) -12 m_t^4 (v_0)
	\right)\right\}.
\eea
In the special case, when $M_0^2=0$, the potential \eqref{EP-v-0K-2} reduces to the Coleman-Weinberg potential.

\section{ ELECTROWEAK PHASE TRANSITION}\label{sec4}

In sequence of SSB of the E331 model, the SSB which breaks the gauge symmetry $ \mathrm{SU}(3)_L \otimes \mathrm{U}(1)_X$
down to the $\mathrm{SU}(2)_L \otimes \mathrm{U}(1)_Y$ through $\chi^0_3$ generates the masses for the exotic quarks,
the heavy gauge bosons $X^0$ and $Z_2$, and gives the first part of mass for $Y^\pm$. The SSB which breaks the symmetry
$\mathrm{SU}(2)_L \otimes \mathrm{U}(1)_Y$ down to the $\mathrm{U}(1)_Q$ through $\chi^0_1$ and $\phi^0_2$ generates
 the masses for the SM particles and gives the last part of mass for $Y^{\pm}$. Because $\om_0 \sim \mathcal{O}(1)$ TeV,
 $u_0 \sim \mathcal{O}(1)$ GeV, and $v_0=246$ GeV \cite{dkck,ecn331}, the breaking $SU(3)\rightarrow SU(2)$ occurs
  before the breaking $SU(2)\rightarrow U(1)$.

Associated with this sequence of SSB, a sequence of EWPT takes place with the transition $SU(3) \rightarrow SU(2)$ at the
scale of $\om_0$ and the transition $SU(2)\rightarrow U(1)$ at the scale of  $v_0$ as the Universe cools down from the hot big bang.
Our analysis so far shows that the former is the first transition which depends only on $\om$, while the latter is the second transition
 which depends only on $v$.

From Table \ref{tab1}, the gauge bosons $X^0$ and $Z_2$ are only involved in the first transition, the gauge bosons
$W^{\pm}$, $Z_1$ and $H^0$ are only involved in the second transition, but the bosons $Y^{\pm}$, $H^0_1$, and
$H^+_2$ are involved in both transitions. The total mass of $Y^{\pm}$ -- i.e. $m_{Y^{\pm}}(\om,v)$, whose formula is
given by \eqref{BosonMassSplit} -- is generated as follows. As the Universe is at the $\om_0$ scale and the EWPT
$SU(3)\rightarrow SU(2)$ happens, $Y^{\pm}$ eats the Goldstone boson $\chi^{\pm}_2$ of the triplet $\chi$ to acquire
 the first part of mass, $m_{Y^{\pm}}(\om)$. When the Universe cools to the $v_0$ scale and the EWPT $SU(2)\rightarrow U(1)$
 is turned on, $Y^{\pm}$ eats the Goldstone boson $\rho^\pm_1$ of triplet $\phi$ and get the last part of mass, $m_{Y^{\pm}}(v)$.

\subsection{Phase transition $SU(3) \rightarrow SU(2)$}

Taking place at the scale of $\om_0$ which is chosen to be in the range $1-5$ TeV, the EWPT $SU(3) \rightarrow SU(2)$
involves exotic quarks and heavy bosons, without the involvement of the SM particles. From Eq. \eqref{EP-o},
 the high-temperature effective potential of the EWPT can be rewritten as
\begin{equation}\label{EP-o-1}
V_{eff}(\om)=D'(T^2-{T'}^2_0){\om}^2-E'T\om^3 +\frac{\lambda'_T}{4}\om^4,
\end{equation}
in which
\bea
D'&=& \frac{1}{24 \om_0^2} \left\{
	6 m_Y^2(\om_0)+6 m_X^2(\om_0)+ 3m_{Z_2}^2(\om_0)
	+m_{H^0_1}^2(\om_0) +2m_{H^+_2}^2(\om_0)
	+18m_Q^2(\om_0)\right\}
	,\crn
{T'}_0^2 &=&  \frac{1}{D'}\left\{
	\frac{1}{4} m_{H^0_1}^2(\om_0) -
 	\frac{1}{32\pi^2\om_0^2} \left(
 		6 m_Y^4(\om_0) +6 m_X^4(\om_0) +3 m_{Z_2}^4(\om_0)
 		\right.\right.	\crn
	&&\qquad \left.\left.
		+m_{H^0_1}^4(\om_0) +2m_{H^+_2}^4(\om_0)-36 m_Q^4(\om_0)
		\right)\right\}	,\crn
E' &=& \frac{1}{12 \pi \om_0^3} (
		6 m_Y^3(\om_0)+6 m_X^3(\om_0)+ 3m_{Z_2}^3(\om_0)
		+m_{H^0_1}^3(\om_0) +2m_{H^+_2}^3(\om_0)
	),\label{D'-T'-E'-lambda'}\\
\lambda'_T &=&
 \frac{m_{H^0_1}^2(\om_0)}{2 \om_0^2} \left\{
 	1 - \frac{1}{8\pi^2 \om_0^2 m_{H_1}^2(\om_0)} \left[
 		6m_Y^4(\om_0) \ln \frac{m_Y^2(\om_0)}{bT^2}
 		+6m_X^4(\om_0) \ln\frac{m_X^2(\om_0)}{bT^2}
 		\right.\right.\crn
&&\qquad \left.\left.
	+3m_{Z_2}^4(\om_0) \ln\frac{m_{Z_2}^2(\om_0)}{bT^2}
	+m_{H^0_1}^4(\om_0) \ln \frac{m_{H_1}^2(\om_0)}{bT^2}
	+2m_{H^+_2}^4(\om_0) \ln \frac{m_{H^+_2}^2(\om_0)}{bT^2}
	\right.\right.\crn
&&\qquad \left.\left.
	-36 m_Q^4(\om_0) \ln\frac{m_Q^2(\om_0)}{b_F T^2}
	\right] \right\},\nn \eea
where $\om_0$ is the value at which the zero-temperature effective potential $V^{0^o K}_{eff}(\om)$ gets the minimum. To acquire $V^{0^o K}_{eff}(\om)$, from $V_{eff}(\om)$ in Eq. \eqref{EP-o} we neglect all terms in the form $F_{\mp}\left(\frac{m}{T}\right)$. The minimum conditions for $V^{0^o K}_{eff}(\om)$ are:

\begin{equation}\label{MinCondition-o}
V^{0^o K}_{eff}(\om_0)=0;\hs
\frac{\partial V^{0^o K}_{eff}(\om)}{\partial \om}\Big|_{\om=\om_0}=0;\hs
 \frac{\partial^2 V^{0^o K}_{eff}(\om)}{\partial \om^2}\Big|_{\om=\om_0}=
 m^2_{H^0_1}(\om) \Big|_{\om=\om_0}.
\end{equation}

From the conditions \eqref{MinCondition-o}, we have the minima of the effective potential \eqref{EP-o-1}:
\be \label{o-0}
\om=0, \quad \om \equiv \om_{c}=\frac{2E'T'_c}{\lambda'_{T'_c}},
\ee
where $\om_{c}$ is a critical VEV of  $\chi$ at the broken state, and $T'_c$ is the critical temperature of phase
 transition which is given by
\be \label{th}
T'_c=\frac{T'_0}{\sqrt{1-E'^2/D'\lambda'_{T'_c}}}.
\ee

\begin{figure}[!ht]
\centering
\includegraphics[height=6cm, width=10cm]{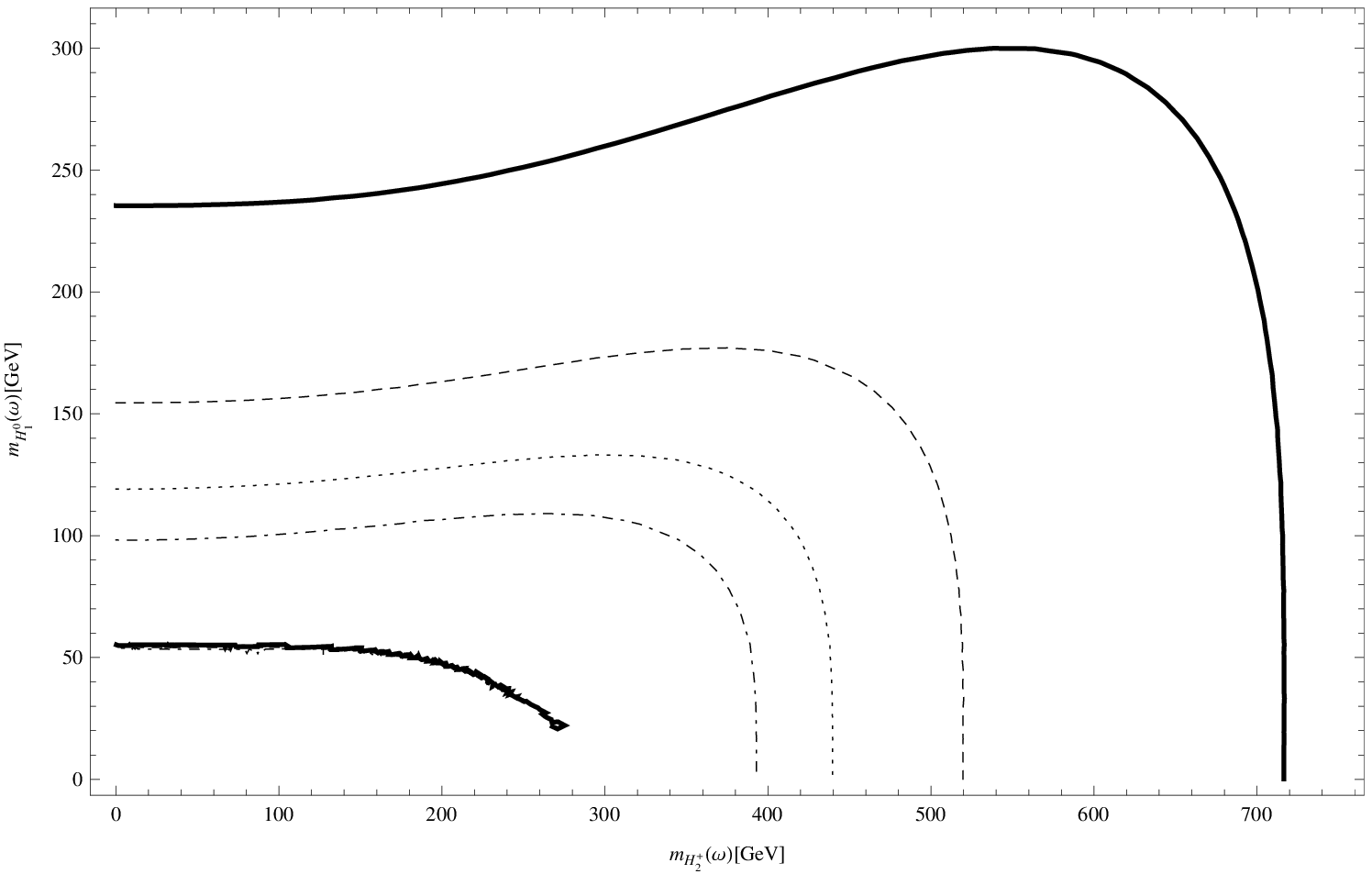}
\caption{The contours of $S'=\frac{\om_c}{T'_c}$ in the case $\om_0=1$ TeV. Solid (and smooth) contour: $S'=1$;
dashed contour: $S'=2$; dotted contour: $S'=3$; dotted-dashed contour: $S'=4$; uneven contour: $S' \to \infty$. In this case,
the mass ranges of $m_{H^0_1}$ and $m_{H^{\pm}_2}$ for the first-order phase transition are $0<m_{H^0_1}<300$
GeV and $0<m_{H^{\pm}_2}<720$ GeV, respectively.}\label{fig:gd1-1}
\end{figure}

\begin{figure}[!ht]
\centering
\includegraphics[height=6cm, width=10cm]{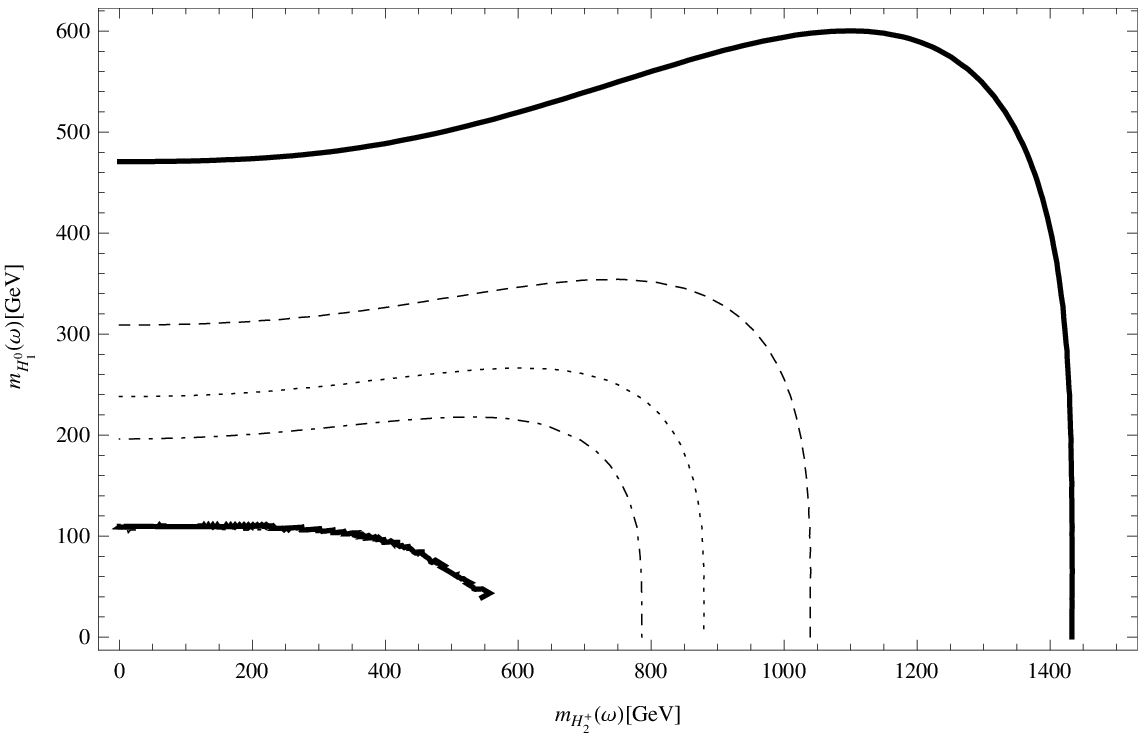}
\caption{The contours of $S'=\frac{\om_c}{T'_c}$ in the case $\om_0=2$ TeV. Solid (and smooth) contour: $S'=1$;
dashed contour: $S'=2$; dotted contour: $S'=3$; dotted-dashed contour: $S'=4$; uneven contour: $S' \to \infty$.
The mass ranges of $m_{H^0_1}$ and $m_{H^{\pm}_2}$ for the first-order phase transition are $0<m_{H^0_1}
<600$ GeV and $0<m_{H^{\pm}_2}<1440$ GeV, respectively.}\label{fig:gd1-2}
\end{figure}

\begin{figure}[!ht]
\centering
\includegraphics[height=6cm, width=10cm]{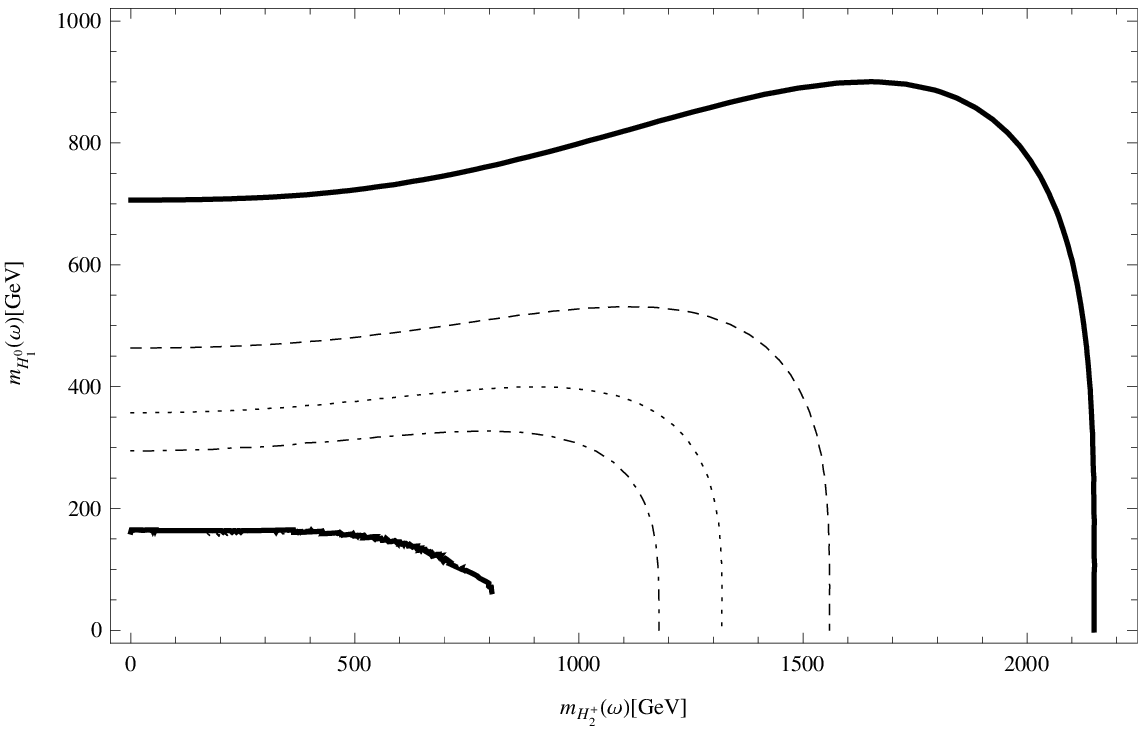}
\caption{The contours of $S'=\frac{\om_c}{T'_c}$ in the case $\om_0=3$ TeV. Solid (and smooth) contour: $S'=1$; dashed
 contour: $S'=2$; dotted contour: $S'=3$; dotted-dashed contour: $S'=4$; uneven contour: $S' \to \infty$. The mass ranges of
 $m_{H^0_1}$ and $m_{H^{\pm}_2}$ for the first-order phase transition are $0<m_{H^0_1}<900$ GeV and
 $0<m_{H^{\pm}_2}<2150$ GeV, respectively.}\label{fig:gd1-3}
\end{figure}

\begin{figure}[!ht]
\centering
\includegraphics[height=6cm, width=10cm]{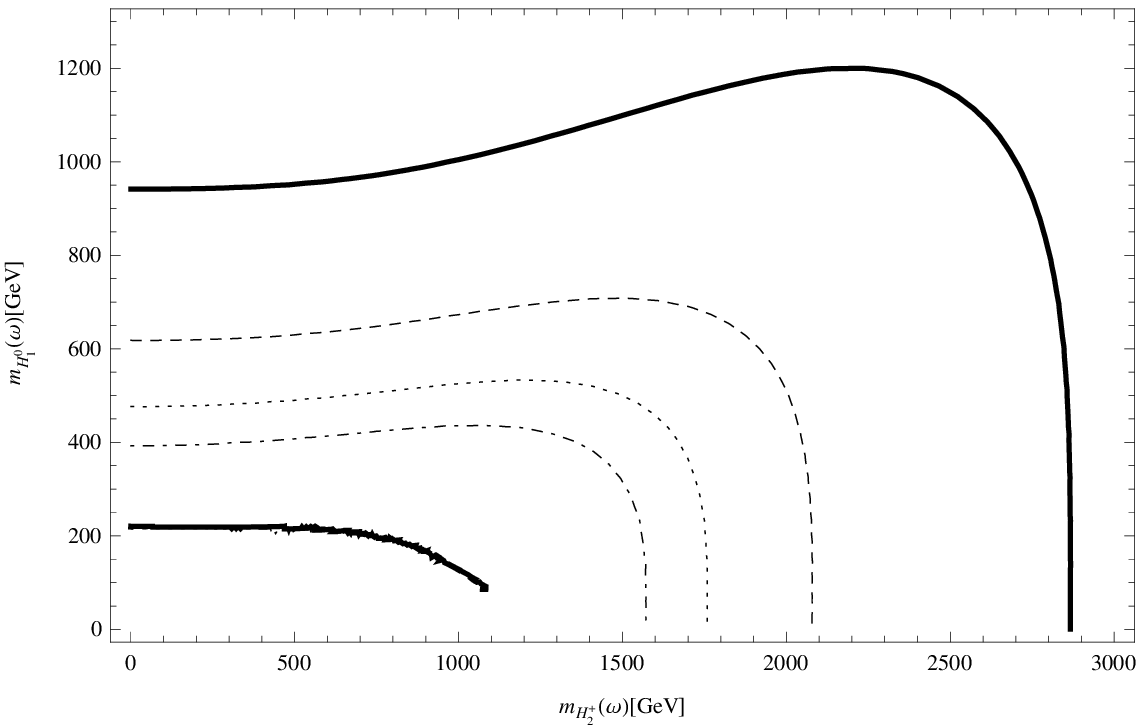}
\caption{The contours of $S'=\frac{\om_c}{T'_c}$ in the case $\om_0=4$ TeV. Solid (and smooth) contour: $S'=1$;
dashed contour: $S'=2$; dotted contour: $S'=3$; dotted-dashed contour: $S'=4$; uneven contour: $S' \to \infty$.
The mass ranges of $m_{H^0_1}$ and $m_{H^{\pm}_2}$ for the first-order phase transition are $0<m_{H^0_1}
<1200$ GeV and $0<m_{H^{\pm}_2}<2870$ GeV, respectively.}\label{fig:gd1-4}
\end{figure}

\begin{figure}[!ht]
\centering
\includegraphics[height=6cm, width=10cm]{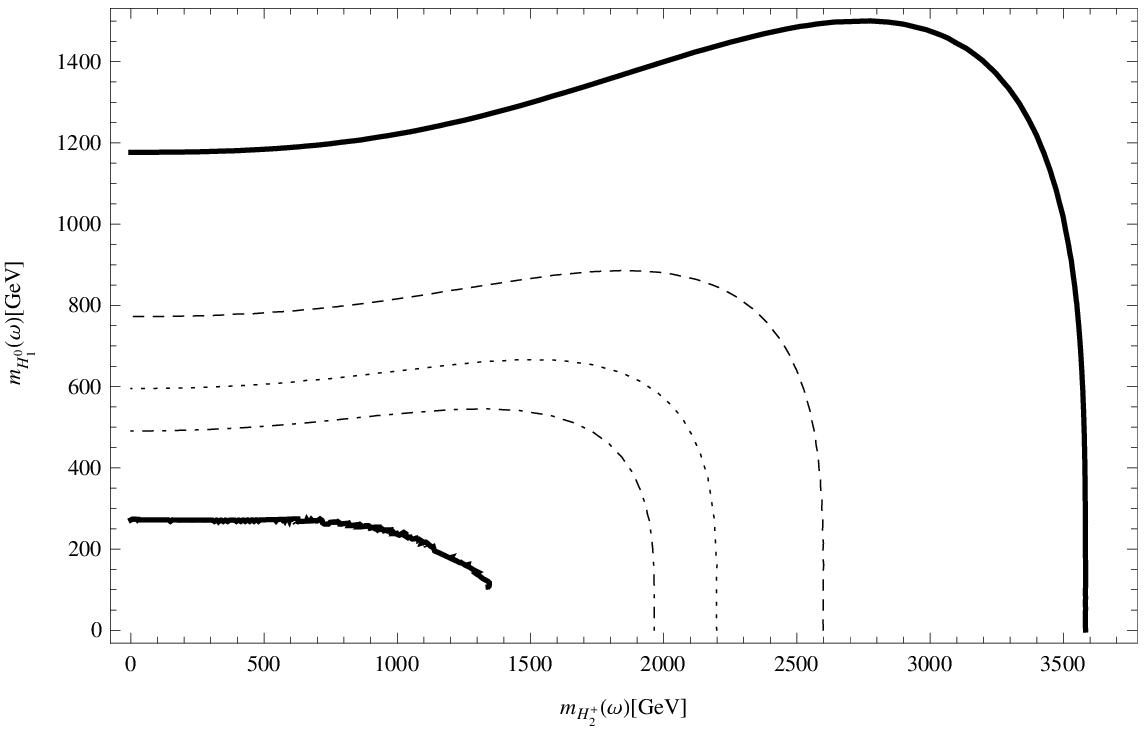}
\caption{The contours of $S'=\frac{\om_c}{T'_c}$ in the case $\om_0=5$ TeV. Solid (and smooth) contour: $S'=1$;
 dashed contour: $S'=2$; dotted contour: $S'=3$; dotted-dashed contour: $S'=4$; uneven contour: $S' \to \infty$.
 The mass ranges of $m_{H^0_1}$ and $m_{H^{\pm}_2}$ for the first-order phase transition are $0<m_{H^0_1}<
 1500$ GeV and $0<m_{H^{\pm}_2}<3590$ GeV, respectively.}\label{fig:gd1-5}
\end{figure}

Now, we consider the phase transition strength:
\be \label{S'}
S'=\frac{\om_c}{T'_c}=\frac{2E'}{\lambda'_{T'_c}},
\ee
which is a function of three unknown masses, $m_{H^0_1}$, $m_{H^{\pm}_2}$ and $m_Q$. For simplicity,
 we follow the ansatz in \cite{ESMCO} and assume $m_{H^{\pm}_2}=m_{Q}$. Then we plot the transition strength
 $S'$ as the function of $m_{H^0_1}(\om_c)$ and $m_{H^{\pm}_2}(\om_c)$ with $\om_c$ is in the range from
 $1$ TeV to $5$ TeV. In Figs. (\ref{fig:gd1-1})-(\ref{fig:gd1-5}), we present the contours of $S'$  in the
 $(m_{H^{\pm}_2}, m_{H^0_1})$-plane; each Fig. corresponds with a case of $\om$. The smooth contours are
  the sets of the $(m_{H^{\pm}_2}, m_{H^0_1})$-pairs which make $S' > 1$ and then the EWPT $SU(3) \rightarrow
  SU(2)$ to be the first-order phase transition. The uneven contours are the sets of the $(m_{H^{\pm}_2}, m_{H^0_1})$
  -pairs which are unusable because they make $S' \to \infty$. Our results show that the heavy particle masses must be in
   the range of a few TeV, and the strength of the first-order phase transition $SU(3) \rightarrow SU(2)$ is in the range $1 < S' < 13$.

According to Ref. \cite{5percent}, the accuracy of a high-temperature expansion for the effective potential such as
 that in Eq. \eqref{EP-o-1} will be better than $5\%$ if $\frac{m_{boson}}{T}< 2.2$, where $m_{boson}$ is the relevant
  boson mass. This requirement sets the "upper bounds" of the mass ranges of $H^0_1(\omega)$ and $H^{\pm}_2(\omega)$.
  From Table \ref{tab2}, this requirement is satisfied by all mass ranges of $H^0_1$, while it narrows slightly most of the
   mass ranges of $H^{\pm}_2$.

\begin{table}
\caption{The mass ranges of $H^0_1$ and $H^{\pm}_2$ for the EWPT $SU(3) \rightarrow SU(2)$ to be the first-order
phase transition, and their upper bounds as required by the condition $m_{boson}< 2.2 \times {T'_c}$.}
\bc
\begin{tabular}{|c|c|c|c|c|}
\hline
\hline
$\om \, [TeV]$ & $T'_c \, [GeV] $ & ${m_{H^0_1}} \, [GeV]$ & ${m_{H^{\pm}_2}} \, [GeV]$ & Upper bound $[GeV]$ \\
\hline
$1$     & $350$   & $0<m_{H^0_1}<300$ & $0<m_{H^{\pm}_2}<720$ &  $770$ \\
\hline
$2$     & $650$   & $0<m_{H^0_1}<600$ & $0<m_{H^{\pm}_2}<1440$ &  $1430$ \\
\hline
$3$     & $950$   & $0<m_{H^0_1}<900$ & $0<m_{H^{\pm}_2}<2150$ &  $2090$ \\
\hline
$4$     & $1300$   & $0<m_{H^0_1}<1200$ & $0<m_{H^{\pm}_2}<2870$ &  $2860$ \\
\hline
$5$     & $1600$   & $0<m_{H^0_1}<1500$ & $0<m_{H^{\pm}_2}<3590$ &  $3520$ \\
\hline
\hline
\end{tabular}
\ec
\label{tab2}
\end{table}

From Eq. \eqref{S'}, the phase transition strength $S'$ depends on the parameters $E'$ and $\lambda'_{T'_c}$.
From Eq. \eqref{D'-T'-E'-lambda'}, $E'$ expresses the contributions of the new bosons while $\lambda'_{T'_c}$
 includes the contributions of the exotic quarks to the phase transition strength. Therefore, the new bosons and exotic
  quarks can be triggers for the EWPT $SU(3) \rightarrow SU(2)$ to be the first-order.

\subsection{Phase transition $SU(2) \rightarrow U(1)$}

Occurring at the scale $v_0=246$ GeV, the phase transition $SU(2) \rightarrow U(1)$ does not involve the exotic quarks or the boson $X^0$. In this stage, the contribution from $Y^\pm$ is equal to that from $ W^{\pm}$. The effective potential is given by Eq. \eqref{EP-v}. We write the high-temperature expansion of this potential as

\begin{equation}\label{EP-v-1}
V_{eff}(v)=D(T^2-T^2_0)v^2-ET|v|^3+\frac{\lambda_T}{4}v^4,
\end{equation}
in which
\bea
D &=&\frac{1}{24 {v_0}^2} \left[
	6 m_W^2(v_0) +6 m_Y^2(v_0) +3m_{Z_1}^2(v_0)
	+m_{H^0}(v_0)+m_{H^0_1}(v_0) +2m_{H^+_2}^2(v_0) +6 m_t^2 (v_0)
 	\right],\crn
T_0^2 &=& \frac{1}{D}\left\{\frac{m_{H}^2(v_0)+m_{H^0_1}(v_0)}{4}
	-\frac{1}{32\pi^2v_0^2} \left(
		6 m_W^4(v_0) +6 m_Y^4(v_0) +3m_{Z_1}^4(v_0)
	\right.\right.\crn
	&&\left.\left.\qquad
		+m_{H^0}^4(v_0)+m_{H^0_1}^4(v_0) +2m_{H^+_2}^4(v_0) -12 m_t^4 (v_0)
	\right)\right\},\crn
E &=& \frac{1}{12 \pi v_0^3} \left(
		6 m_W^2(v_0) +6 m_Y^3(v_0) +3m_{Z_1}^3(v_0)
		+m_{H^0}^3(v_0) +m_{H^0_1}^3(v_0) +2m_{H^+_2}^3(v_0)
	\right),\label{D-T-E-lambda}\\
\lambda_T &=&
 \frac{m_{H^0}^2(v_0)+m_{H^0_1}^2(v_0)}{2 v_0^2}\left\{
 	1- \frac{1}{8\pi^2 v_0^2 (m_{H^0}^2(v_0)+m_{H^0_1}^2(v_0))}\left[
 		6 m_W^4(v_0) \ln \frac{m_W^2(v_0)}{b T^2} 		
 	\right.\right.\crn
	&&\qquad \left.\left.
+6 m_Y^4(v_0) \ln \frac{m_Y^2(v_0)}{b T^2}		
+3m_{Z_1}^4(v_0) \ln\frac{m_{Z_1}^2(v_0)}{b T^2}+m_{H^0}^4(v_0) \ln \frac{m_{H^0}^2(v_0)}{b T^2}		
		\right.\right.\crn
&&\qquad  \qquad  \qquad  \qquad \left.\left.
		+2m_{H^+_2}^4(v_0)\ln\frac{m_{H^+_2}^2(v_0)}{bT^2} 		
		-12 m_t^4(v_0)\ln\frac{m_t^2(v_0)}{b_F T^2}
	\right] \right\},\nn \eea
where $v_0$ is the value at which the zero-temperature effective potential $V^{0^o K}_{eff}(v)$ gets the minimum. Here, we acquire $V^{0^o K}_{eff}(v)$ from $V_{eff}(v)$ in Eq. \eqref{EP-v} by neglecting all terms in the form $F_{\mp}\left(\frac{m}{T}\right)$. 

From the minimum conditions for $V^{0^o K}_{eff}(v)$ 
\begin{equation}\label{MinCondition-v}
V^{0^o K}_{eff}(v_0)=0,\hs
\frac{\partial V^{0^o K}_{eff}(v)}{\partial v}\Big|_{v=v_0}=0,\hs
 \frac{\partial^2 V^{0^o K}_{eff}(v)}{\partial v^2}\Big|_{v=v_0}=
 \left[m^2_{H^0}(v)+m^2_{H^0_1}(v)\right] \Big|_{v=v_0},
\end{equation}
we can see that in this EWPT, $m^2_{H^0}(v)$ and $m^2_{H^0_1}(v)$ generate the masses of the SM particles
and the last mass part of $Y^\pm$. We also have the minima of the effective potential \eqref{EP-v-1}:
\be \label{v-0}
v=0, \quad v \equiv v_c=\frac{2ET_c}{\lambda_{T_c}},
\ee
where $v_c$ is the critical VEV of  $\phi$ at the broken state, and $T_c$ is the critical temperature of phase transition which is given by
\be \label{Tc}
T_c=\frac{T_0}{\sqrt{1-E^2/D\lambda_{T_c}}}.
\ee 

\begin{figure}[!ht]
\centering
\includegraphics[height=6cm, width=10cm]{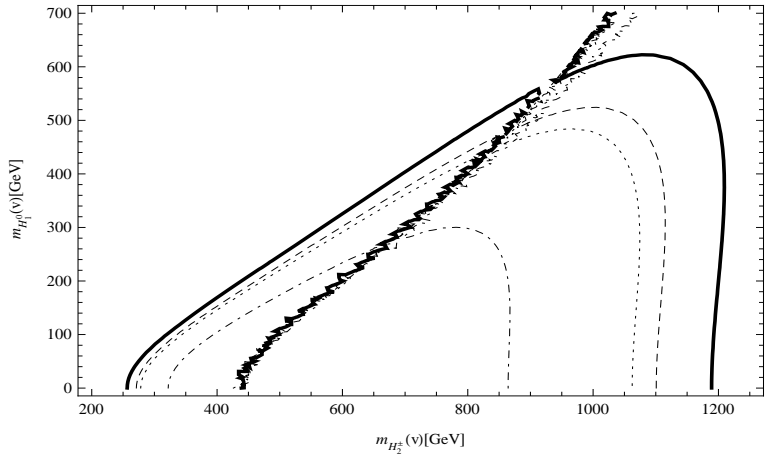}
\caption{The contours of transition strength $S=\frac{2E}{\lambda_{T_c}}$. Solid smooth contour: $S=1$;
dashed smooth contour: $S=1.1$; dotted smooth contour: $S=1.15$; dash-dotted smooth contour: $S=1.5$;
even contours: $S \to \infty$. The mass ranges of $m_{H^0_1}$ and $m_{H^{\pm}_2}$ for the EWPT $SU(2)
 \rightarrow U(1)$ to be the first-order are $0 \, \mathrm{GeV} <m_{H^{0}_1(v)}<620 \,\mathrm{GeV}$
  and $250 \, \mathrm{GeV} <m_{H^{\pm}_2(v)}<1200\,\mathrm{GeV}$, respectively.}\label{fig:8}
\end{figure}

We investigate the phase transition strength
\be \label{S}
S=\frac{v_{c}}{T_c}=\frac{2E}{\lambda_{T_c}}
\ee
of this EWPT. In the limit $E \rightarrow 0$, the transition strength $S \rightarrow 0$ and the phase transition
 is a second-order. To have a first-order phase transition, we requires $S \geq 1$. We plot $S$ as
 \texttt{a}  function of
 $m_{H^0_1}(v_0)$ and $m_{H^{\pm}_2}(v_0)$. As shown in Fig. \ref{fig:8}, for the masses of $H^{\pm}_2$
  and $H^{0}_1$ which are respectively in the ranges $250 \,  \mathrm{GeV} <m_{H^{\pm}_2(v)}<1200 \,
  \mathrm{GeV}$ and $0 \, \mathrm{GeV} < m_{H^{0}_1(v)} < 620 \, \mathrm{GeV}$, the transition
   strength is in the range $1 \leq S <3$.

\begin{figure}[!ht]
\centering
\includegraphics[height=8cm, width=10cm]{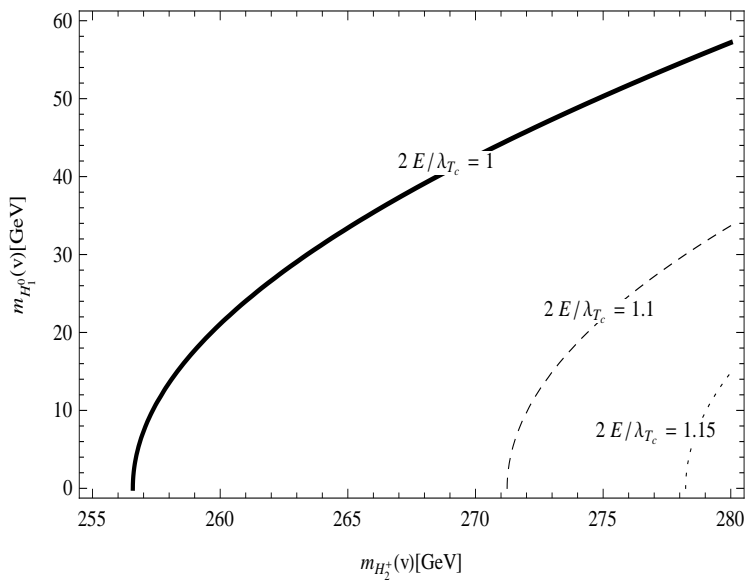}
\caption{The condition $\frac{m_{boson}}{T}< 2.2$ narrows the mass ranges of  $H^\pm_2$ and $H^0_1$ as
 well as the range of transition strength.}\label{fig:11}
\end{figure}

Considering the requirement for the high-temperature expansion to be applicable on the effective
 potential \eqref{EP-v}, $\frac{m_{boson}}{T}< 2.2$  \cite{5percent}, we show in Fig. \ref{fig:11} that
  with $T=T_c \sim 130 \, \mathrm{GeV}$, the mass ranges of  $H^{\pm}_2$ and $H^0_1$ are respectively narrowed to:
\begin{equation} \label{MR-H2-v}
255 \, \mathrm{GeV}<m_{H^{\pm}_2}<280 \, \mathrm{GeV},
\end{equation}
and
\begin{equation} \label{MR-H1-v}
0 \, \mathrm{GeV} <m_{H^{0}_1} < 58 \, \mathrm{GeV}.
\end{equation}

Corresponding with these ranges of mass, the range of phase-transition strength is narrowed to $1 \leq S <1.15$.
 Thus the EWPT $SU(2) \rightarrow U(1)$ is the first-order phase transition, but it seems quite weak.

As we can see in Eqs. \eqref{S} and \eqref{D-T-E-lambda}, the new bosons contribute to the phase transition
 strength $S$ via the parameters $E$ and $\lambda_{T_c}$. Hence these new bosons can be triggers for the EWPT
 $SU(2) \rightarrow U(1)$ to be the first-order.

\begin{figure}[!ht]
\centering
\includegraphics[height=12cm, width=10cm]{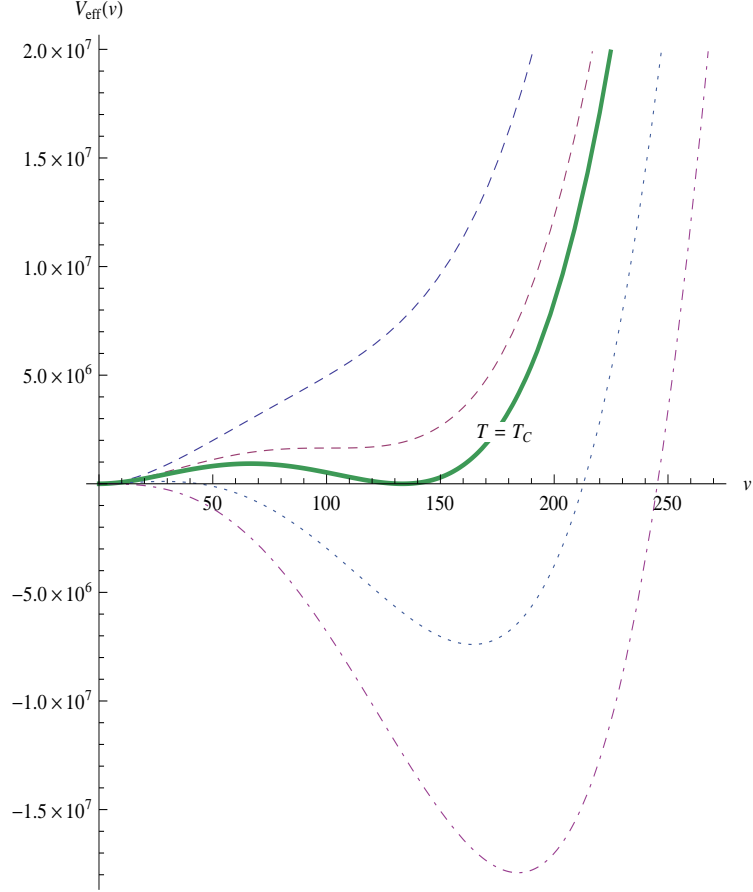}
\caption{The dependence of the effective potential $V_{eff}(v)$ on the temperature. With $m_{H^0_1}(v)=50 \,
 \mathrm{GeV}$ and $m_{H^\pm_2}(v)=280 \, \mathrm{GeV}$, we have the critical temperature $T_c=127.974 \,
 \mathrm{GeV}$ and the phase-transition strength $S=1.03$. Solid line: $T_c$; lines above the solid line: $T>T_c$; lines
  under the solid line: $T<T_c$.
}\label{fig:9}
\end{figure}

\begin{figure}[!ht]
\centering
\includegraphics[height=8cm, width=10cm]{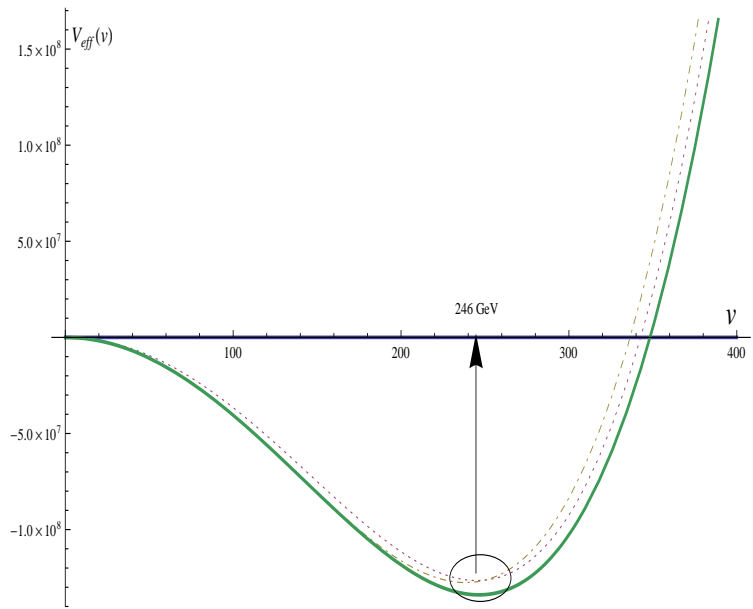}
\caption{The tendency of nonzero minimum for lower temperatures. We choose $m_{H^{0}_1}(v)=50 \, \mathrm{GeV}$, $m_{H^{\pm}_2}(v)=280 \, \mathrm{GeV}$. Dot-dashed line: $T=50 \, \mathrm{GeV}$. Dotted line: $T=10 \, \mathrm{GeV}$. Solid line: $T=1 \,
\mathrm{GeV}$. $v_0$ reaches to $246 \, \mathrm{GeV}$ as the temperature decreases.}\label{fig:10}
\end{figure}

In Fig. \ref{fig:9}, we illustrate the dependence of the effective potential $V_{eff}(v)$ on the temperature. When the
Universe cools through the phase-transition critical temperature $T_c$, the Higgs field $v$ tends to get a nonzero VEV
$v_0$ which is in the range $0<v_0<246 \, \mathrm{GeV}$, and the second minimum of $V_{eff}(v)$ gradually
 appears at $v_0$. As the temperature drops from $T_c$, the second minimum becomes lower and the first
  minimum gradually disappears, while the VEV $v_0$ tends to $246 \, \mathrm{GeV}$. The tendency of $v_0$
  can be seen in Fig. (\ref{fig:10}) where we show that $v_0$ reaches to $246 \, \mathrm{GeV}$ for the temperatures
  which are far below $T_c$. At $0^o K$, the non-zero minimum locates exactly at $v_0 =246$ GeV. This result is consistent with the SM.

\subsection{Constraint on the mass of the charged Higgs boson }
From the EWPT $SU(2)\rightarrow U(1)$, we have derived the mass ranges of ${H^+_2}(v)$ and ${H^0_1}(v)$ in
 Eqs. \eqref{MR-H2-v} and \eqref{MR-H1-v}. So we have
\begin{equation}\label{M-H1}
0 \,\mathrm{GeV}  < m_{H^0_1}=\sqrt{m^2_{H^0_1}(v)+m^2_{H^0_1}(\omega)} < 1501.12 \,\mathrm{GeV},
\end{equation}
and we obtain
\be
2.149< \lambda_4 < 2.591, \label{hq1}
\ee
and
\be
0<\frac{\lambda^2_3}{2\lambda_1}<0.0556,
\ee

From the phase transition $SU(3)\rightarrow SU(2)$, we have also derived
\be
0< \lambda_4 < 10.3, \label{hq2}
\ee
and
\be
0< \lambda_1< 0.45, \label{hq3}
\ee
for any $\omega$. Eqs. (\ref{hq1})-(\ref{hq3}) lead to $ 2.149< \lambda_4 < 2.591$; $0< \lambda_1< 0.45$ and
$0<\frac{\lambda^2_3}{2\lambda_1}<0.0556$.

\section{CONCLUSION AND OUTLOOKS}\label{sec5}

We have investigated the EWPT in the E331 model using the high-temperature effective potential. Although the
effective potential in the model depends complicatedly on three VEVs, $u$, $\om$, and $v$, it can be transformed
 to a sum of two parts so that each part depends only on $\om$ or $v$, which corresponds a stage of SSB. Thanks to that
  the EWPT can be seen as a sequence of two EWPTs. The first, $SU (3)\rightarrow SU(2)$, takes place at the energy scale
  $\om_0$ to generate the masses for the exotic quarks, the heavy gauge bosons $X^0$ and $Z_2$, as well as a mass part
  of $Y^\pm$. The second, $SU(2)\rightarrow U(1)$, occurs at the scale $v_0$ to give the masses for the SM particles and
  the remained mass part of $Y^{\pm}$.

At the TeV scale, the EWPT $SU(3) \rightarrow SU(2)$ is strengthened by the new bosons and the exotic quarks to
be the strongly first-order; if the masses of these new particles are about $10^2 - 10^3$ GeV, the phase transition strength
is in the range $1 - 13$. As the energy is lowered to the scale of $10^2$ GeV, the EWPT $SU(2) \rightarrow SU(1)$
is strengthened by only the new bosons; with the contributions of the mass parts from $H^0_1$, $H^\pm_2$ and $Y^\pm$
 which are in the ranges $10 - 10^2$ GeV, the strength of this transition is about $1 - 1.15$. Therefore, both EWPTs can be
  the first-order; the $SU(3) \rightarrow SU(2)$ appears very strong, while the $SU(2) \rightarrow SU(1)$ seems quite weak.

However, both of these first-order EWPTs can be sufficiently strong to provide B violation necessary for baryogenesis,
 as shown via the parameter ranges which we have specified. If $H^0_1$ and $H^{\pm}_2$ exist, their contributions to the
 strengths of each EWPT are meaningly large. In this case, the sequence of strongly first-order EWPTs in the model may
  provide a source of large deviations from thermal equilibrium. And the model may fully describe the continual existence
  of BAU since being generated in the early Universe.

In the next works, we will investigate the electroweak sphalerons as well as the C- and CP- violating interactions
 to know if the model possesses all necessary components for EWBG.

\section*{ACKNOWLEDGMENTS}
This research is funded by Vietnam  National Foundation for Science and Technology Development (NAFOSTED)  under grant number
103.01-2014.51.
\\[0.3cm]

\end{document}